# Degeneracy along the sensorimotor hierarchy: motor control within a framework larger than redundancy

PACLET Florent [1] and DUPRAT Paul [1]

[1] Institut de Neurosciences Cognitives et Intégratives d'Aquitaine, UMR 5287, CNRS, Université de Bordeaux, Bordeaux, France

**Summary :** Motor control has described the surplus of solutions available to the nervous system as redundancy, a term that names duplication: interchangeable elements, robust to loss but incapable of differential adaptation. Biology has had a second term for twenty-five years. Degeneracy names elements that are not interchangeable and are nonetheless isofunctional with respect to a given output, and it supports adaptability, since non-identical elements necessarily diverge in some context. Circuit neuroscience has relabeled its own results accordingly, while motor control has kept the older vocabulary. Neuromechanical models, by placing a spinal circuit in the loop with a musculoskeletal apparatus, bring the two traditions onto the same class of objects. We restate the Edelman and Tononi distinction for sensorimotor systems and derive an operational requirement: not the existence of multiple solutions, but their divergence in contexts they were not selected for. Three influential studies each meet part of that requirement and none meets all. We then argue that degeneracy and redundancy coexist along the sensorimotor hierarchy in a proportion that varies continuously, and that this proportion is measurable: computing degeneracy twice for the same configuration, once with muscle activation as the output and once with the movement produced, isolates what the musculoskeletal apparatus contributes. Five predictions follow, with the single outcome that would refute the proposal. We set out the adaptations the measurement requires in a nonlinear, non-stationary, closed-loop system, and what changes for motor control once solutions are no longer assumed equivalent: the question shifts from which rule selects a command to what the repertoire of the system still allows.

## 1. Introduction

When Bernstein asked skilled blacksmiths to strike the same point repeatedly with a hammer, the kinematics of the arm differed on every trial while the point of impact stayed invariant (Bernstein, 1967). The observation set one of the founding problems of motor control: the body has more degrees of freedom than any task requires, so that infinitely many joint configurations, muscle activations, and neural commands can produce the same outcome. The uncontrolled manifold hypothesis showed that this variability is selectively channeled into the dimensions that do not matter for the task (UCM; Scholz and Schöner, 1999), and Latash proposed abundance rather than redundancy, the surplus of effectors being a resource the system exploits rather than a problem it must solve (Latash, 2012). Both frameworks address the same object, the mapping from effector states to task outcomes, examined at the periphery through forces. In that tradition, redundancy is a property of the musculoskeletal system.

The modeling response followed a path of increasing constraint. Early models reduced the limb to rigid segments driven by lumped torque generators (Seireg and Arvikar, 1973; Crowninshield and Brand, 1981), in which redundancy is minimal by construction: with one actuator per degree of freedom, the mapping is invertible and inverse dynamics is enough. Adding realistic muscles brings the problem back, since Hill-type musculotendon units outnumber the joint degrees of freedom and the projection from muscle forces to joint moments is no longer invertible (Zajac, 1989). The standard resolution has been static optimization: among all activation patterns compatible with the observed movement, the one that minimizes a cost is retained (Crowninshield and Brand, 1981; Todorov and Jordan, 2002). The choice of cost determines the solution, and different costs return different solutions from identical kinematics. Later developments mostly added constraints, up to replacing the optimization with measured excitations (Buchanan et al., 2004) or characterizing the feasible activation space rather than a single point within it (Valero-Cuevas, 2015). Another class of constraint comes not from the mechanics of the limb but from control itself. Hand models long treated each finger as mechanically independent, an assumption at odds with the anatomical and neural coupling between fingers (Li et al., 1998; Zatsiorsky et al., 2000; Danion et al., 2003). Imposing wrist equilibrium substantially changes the tendon tensions estimated from the same data, and the resulting co-contraction pattern accounts for the force deficit reported in pressing tasks (Paclet and Quaine, 2012). The decisive constraint was an assumption about control, not an anatomical fact. These advances shrink the solution space without removing the indeterminacy, which is inherent to the system. What was still missing was an explicit controller, and neuromechanical models supply it. Rather than selecting among musculoskeletal solutions with an external cost, they embed a neural circuit that generates the activations, and they close the loop through proprioceptive feedback (Raphael et al., 2010; Parziale et al., 2020; Cattaert et al., 2026). The musculoskeletal system is no longer something to invert but one half of a coupled system: the circuit produces activations, the limb moves, and the movement returns as afferent input. The multiplicity of solutions does not disappear, it moves into the circuit. The α motoneuron is a terminal effector, whose discharge sets muscle activation, and also the convergence point of multiple pathways of heterogeneous origin: descending commands, mono- and polysynaptic reflex arcs, Ia and Ib afferents, Renshaw and propriospinal circuits. A model that reproduces this architecture therefore inherits a command space far larger than the behavior map onto which it projects.

The consequences have been established at scale in central pattern generators (CPGs). A screen of twenty million versions of the crustacean pyloric circuit showed that parameter sets spanning two orders of magnitude in synaptic conductance produce the same rhythm (Prinz et al., 2004). Read through the vocabulary inherited from motor control, this is neural redundancy: more parameters than the task requires, hence many solutions. The description is accurate, and it establishes how many solutions exist and how far apart they lie. It does not establish whether those solutions are equivalent, and the two possibilities have opposite consequences. If the solutions are interchangeable, the system is robust, but any perturbation that affects one affects all of them. If they are not, the system holds a repertoire within which a perturbation can eliminate some solutions while sparing others. That this distinction has remained open is a documented concern. Neuroscience often proceeds as if its measurements obeyed a single function $y = f(x)$, whereas the relation is better written $y = f_i(x)$, where $i$ indexes the individual and its own parameter set; averaging across individuals then reduces distinct configurations to a mean that may match none of them (Bernard, 2023).

The concept behind this concern was introduced in biology a quarter of a century ago. **Redundancy** in its technical sense means duplication: the same function carried by interchangeable elements, that is, elements none of which does anything the others do not. They are robust to loss but rigid, since being equivalent in every context they cannot support any differential adaptation. **Degeneracy** refers to elements that are not interchangeable and are nonetheless isofunctional with respect to a given output (Edelman, 1987; Tononi et al., 1999; Edelman and Gally, 2001). Not being identical, there are necessarily contexts in which they diverge, which identical elements cannot do. Degeneracy is therefore the substrate of adaptability. This divergence has been demonstrated directly in experiment: two circuits of the crab gastric mill, isofunctional under baseline conditions, separate under one perturbation and remain indistinguishable under another (Powell et al., 2021). Strictly redundant elements cannot behave this way, and that is exactly the information a count leaves out. Tononi and colleagues provided information-theoretic measures that make both concepts quantifiable with respect to a defined set of outputs, and that establish them as nested rather than opposed. A degenerate system necessarily shows some overlap between contributions; a fully redundant system is not degenerate, because the independent effects of its elements have been lost. Since degeneracy is formalized as a function of the overlap measured over subsets of increasing size, the approach developed here does not aim to refute the established literature on redundancy but to place it within a broader theoretical frame. In circuit neuroscience, the relabeling is documented within the output of a single laboratory: results reported in 2004 as disparate parameters were described as degeneracy by 2021, with explicit reference to Edelman (Prinz et al., 2004; Powell et al., 2021). The framework has since been consolidated at scales ranging from bacteria to human cognition (Goaillard and Marder, 2021; Albantakis et al., 2024). In machine learning, solution degeneracy is quantified across trained recurrent networks, behavioral degeneracy being operationalized as the variability of responses to out-of-distribution inputs (Huang et al., 2025).

Motor control has largely kept the term redundancy. The reason is historical rather than conceptual: the problem was posed at the periphery, where the term is often appropriate, and the vocabulary traveled with the models toward the higher levels. As long as those models remained musculoskeletal, little was lost. Neuromechanical models, by placing a spinal circuit in the loop, move motor control onto the ground that circuit neuroscience has described as degenerate for two decades: the two traditions now converge on the same class of objects and describe it in incompatible terms. This includes our own contribution, since Cattaert et al. (2026) attributed to the redundancy of parallel sensorimotor pathways a diversity of descending commands that meets the requirement for degeneracy. This is an illustration rather than an isolated oversight: where no formal framework separates the two properties, cases of degeneracy are folded into redundancy.

Yet the nervous system and the musculoskeletal apparatus are not of the same nature. The central nervous system is heterogeneous, shaped by selection, and adaptive on short timescales; the musculoskeletal apparatus is mechanical, constrained by the physics of levers and contraction, and treated as fixed at our modeling scale. One would therefore expect degeneracy to characterize the first and redundancy the second. We will show that this expectation does not hold, and propose that the two properties coexist rather than succeed one another, in proportions that vary continuously along the sensorimotor hierarchy and that depend on the output considered, on the type of perturbation applied, and on the timescale over which it acts. Stated this way, the hypothesis is measurable: changing the definition of the output moves the boundary of the observed system, and the gap between the two measurements quantifies what the periphery contributes to the degeneracy of the whole.

Establishing degeneracy empirically requires three conditions rarely met together in experimental preparations: (I) identifying several solutions that produce the same behavior, (II) characterizing their internal organization, (III) testing them in contexts they were not selected for. Neuromechanical models meet all three: the state of every neuron and every synapse is accessible at every instant, lesions are applied with arbitrary precision and full reversibility, and exploration algorithms build the behavior map far more efficiently than exhaustive sampling (Forestier et al., 2017). Our examples draw on one such model, a spinal circuit controlling an elbow joint driven by two antagonist muscles (Cattaert et al., 2026), but the argument holds for any system in which several command sets produce the same behavior. Its single degree of freedom removes kinematic redundancy, since one joint admits only one configuration for a given angle, while leaving muscular redundancy in the antagonist pair, in the form of co-contraction. The system is thus placed in minimal contact with mechanics, which makes their coupling easier to observe.

This article is deliberately conceptual. Its aim is not to report new measurements but to give motor control the means to distinguish two properties that the available vocabulary does not separate, and to measure them. We first restate the distinction of Edelman and Tononi in terms that apply to sensorimotor systems, and draw from it an operational requirement: not the existence of multiple solutions, but their divergence in contexts they were not selected for. We then hold three studies from the literature against that requirement, to test whether any of them meets it in full. We then propose that degeneracy and redundancy coexist along the sensorimotor hierarchy, in a proportion that the gap between two definitions of the output makes it possible to quantify. From this we draw the predictions that would test the proposal and the result that would refute it, then the conditions under which the measurement can be carried out, before examining what it implies for motor control.

## 2. The Edelman and Tononi framework restated for sensorimotor systems

Degeneracy was originally defined as the property by which non-isomorphic structures are isofunctional with respect to a given output (Edelman, 1987). The definition leaves two terms open, and both have to be settled before any use. The first is non-isomorphic. A structural reading applies easily in engineering but poorly in biology (Tononi et al., 1999): no two motoneuron pools are identical, and no two parameter sets are either, since they differ numerically by construction. Taken structurally, almost everything in a biological system is non-isomorphic, and the condition loses its discriminating power. The second term is the output. Isofunctionality is never absolute: two spinal configurations producing flexions of 43° and 47° are isofunctional if the output retained is amplitude, and cease to be if it is the time course of motoneuron discharge. The formalism that follows from the definition says how to compute once the output is fixed, not how to fix it. For an artificial network with identified output units, the question barely arises. For a sensorimotor system, where the movement produced returns through the afferents, it arises in full, and we will show that it has a useful answer. Tononi and colleagues resolved both difficulties by moving to the functional level, defining degeneracy and redundancy in information-theoretic terms with respect to a given set of outputs. What counts is no longer an absolute structural distinction between two components, but the capacity of each to influence the output in ways that are at once redundant and functionally separable.

The structural reading has not disappeared, and it still decides verdicts. A recent review illustrates the distinction on five neurons (Albantakis et al., 2024). In the first case, a neuron receives four suprathreshold inputs: each is sufficient on its own to make it discharge, none differs from the others, and the authors treat them as redundant causes. In the second, the same neuron receives four subthreshold inputs of different weights: only certain combinations make it discharge, and those combinations are called degenerate. The input neurons did not change from one case to the other, only the weight of their synapses did. It is therefore that weight which counts as part of the structure, without the choice being discussed, and restricting the structure to the cell body would move the second case back to redundancy. The output is thus not the only term left open by Edelman's definition: the grain at which the system is described bears on the verdict as well. The same illustration acknowledges that the attribution holds for a given context, since the effects may differ if the discharge threshold moves.

A word on vocabulary, since several usages circulate. The structural view defines redundancy as the presence of identical copies. We adopt a strictly functional one. Functionally, whether the underlying elements are identical or different is beside the point; what counts is the overlap of their effects on the output. Both views have their use, and the structural one supports anatomical inference. But quantifying shared information requires isolating the functional effect, so our elements are evaluated by their mapping to the output and by nothing else. Cellular neurophysiology has named functional overlap the case where two elements share a function while each keeps effects the other does not have, and places there most of the cases described so far as redundant or degenerate (Goaillard and Marder, 2021); this is exactly the configuration the measure presented below is built to isolate. We therefore keep to the pair redundancy and degeneracy, without subdividing them: neither exists in a structural version and a functional version.

One consequence follows, and it governs everything below. Observing a system in a single context does not separate the two properties: the same behavioral response remains compatible with either. What separates them is their counterfactual behavior. The interchangeability of redundant elements couples their successes and their failures strictly, whereas degenerate elements keep autonomous responses to novel conditions. That autonomy rests on a biological inference, which physiological observations of circuits diverging under strain support: elements that are not structurally identical do not respond identically once conditions change. Environmental or task perturbation is therefore not a methodological convenience, it is constitutive of the distinction: a system can be shown to be degenerate only by moving it away from the conditions under which its isofunctionality was established. For sensorimotor systems the implication is direct: cataloguing the solutions a model can produce, however exhaustively, establishes their multiplicity and nothing more.

Everything rests on a single quantity, mutual information. It measures what two things have in common: if knowing one reduces the uncertainty about the other, they share information. Applied here, it compares a subset of neurons, written $X_j^k$ (the $j$-th subset of $k$ elements drawn from a network of $n$), and an output $O$:

$$\mathrm{MI}(X_j^k\,;\,O) = H(X_j^k) + H(O) - H(X_j^k, O) \qquad (1)$$

where $H$ is entropy, the uncertainty about the state of a variable. The first two terms are the uncertainties of $X_j^k$ and of $O$ taken separately, the third that of the pair taken as a whole. Decomposing this last term, the equation can be rewritten $\mathrm{MI}(X_j^k\,;\,O) = H(O) - H(O \mid X_j^k)$: mutual information measures the reduction in uncertainty about the output gained from knowing that subset, and conversely.

This measure has one decisive limitation for our object: it is symmetric, hence blind to the direction of causation. It indicates that two variables are related, without saying which acts on which. Three very different situations produce the same value, and in a spinal sensorimotor circuit all three are present at once by construction. Motoneurons command muscle activation: the subset drives the output. Muscle length, velocity, and tension are picked up by spindles and tendon organs, then returned to that same circuit: the output feeds back on the subset. And meanwhile, descending commands set the excitability of interneurons while triggering the movement that feeds those afferents: a third element drives both the subset and the output. Observing the correlations of such a system therefore does not yield its causes.

While task perturbations reveal counterfactual behavior, isolating causality in closed loops requires a second, probing perturbation. Tononi and colleagues answer this by applying a targeted causal perturbation rather than mere observation: uncorrelated noise of fixed variance is injected into each subset in turn, and mutual information is evaluated on the perturbed system. The quantity obtained for each perturbed subset, $\mathrm{MI}^{\mathrm{P}}$, keeps the form of equation (1) but acquires a causal interpretation, since any change in the output can only be attributed to the perturbed subset. Neuromechanical models lend themselves to this remarkably well: the noise can be applied selectively, removed without residual effect, and the same parameter set replayed under altered conditions, which is impossible in a biological preparation.

Redundancy is the excess of the sum of individual contributions ($k$ = 1) over that of the system taken as a whole:

$$R(X\,;O) = \Sigma_j\, \mathrm{MI^P}(X_j^1\,;O) - \mathrm{MI^P}(X\,;O) \qquad (2)$$

It is zero when each element contributes independently to the output, the sum of individual contributions then equaling the global information, and high when each element taken alone already carries most of what the whole carries. This reading calls for one precaution. The second term of the equation covers both redundancy and synergy, the part that appears only when several elements are taken together (Barrett, 2015), so the subtraction actually computes a net balance, redundancy minus synergy. A zero term does not necessarily mean the absence of redundancy, it may reflect strict compensation between the two.

Degeneracy is defined as the departure from proportionality with subset size:

$$D_{\mathrm{N}}(X\,;O) = \Sigma_k\, [\, \langle \mathrm{MI^P}(X_j^k\,;O)\rangle - (k/n)\, \mathrm{MI^P}(X\,;O)\, ] \qquad (3)$$

where $\langle\cdot\rangle$ is the average over all subsets of size $k$. The second term is the linear reference: if subsets contributed in proportion to their size, a subset of $k$ elements would supply a fraction $k/n$ of the total information. Degeneracy measures the cumulated departure from that line. The relation between the two quantities becomes explicit when equation (3) is rewritten in terms of redundancy evaluated at each scale:

$$D_{\mathrm{N}}(X\,;O) = \Sigma_k\, [\, (k/n)\, R(X\,;O) - \langle R(X_j^k\,;O)\rangle\, ] \qquad (4)$$

This form has to be read at two scales. Take the whole system first. Its contribution overlap is high: several of its elements carry the same information about the output, and some can be removed without changing the result. The first term sets a theoretical reference, the fraction $k/n$ of that whole-system redundancy, which is the share of overlap expected at a given size if overlap were proportional. Now take small subsets, a few elements at a time. If the elements were simple copies of one another, the overlap of small groups would approach that reference, and overlap would be proportional at every scale. That is what the second term evaluates: is the actual overlap within small groups as strong as it would be for exact copies, or is it weaker? The equation being a subtraction, degeneracy is high when the answer is weaker. Seen from afar, the elements do the same thing; seen up close, each keeps a share of its own. Neither term suffices on its own, and their combinations define three distinct regimes (Table 1).

Table 1: Independence, redundancy, and degeneracy according to the behavior of the two terms of equation (4). The overlap of the whole system and the overlap of small subsets are read in the first and the second term respectively; the gap between them defines degeneracy.

| | **Independence** | **Redundancy** | **Degeneracy** |
|---|---|---|---|
| Contribution overlap | none | high | high |
| Independent effects | high | none | preserved |
| $R$, whole system | ≈ 0 | high | high |
| $\langle R \rangle$, small subsets | ≈ 0 | high | low |
| $\boldsymbol{D_N(X\,;O)}$ | **≈ 0** | **low** | **high** |
| Under perturbation | each loss removes a function | elements fail together | elements diverge |

Two consequences follow. The two properties are nested rather than opposed: a degenerate system necessarily shows contribution overlap, since the first term of equation (4) requires it, whereas a fully redundant system is not degenerate, the second term offsetting the first entirely. Overlap is a necessary but not sufficient condition for degeneracy.

Equation (4) also expresses degeneracy entirely in terms of overlap evaluated over subsets of increasing size. Measuring that overlap at a single scale is therefore not an alternative to measuring degeneracy, but the evaluation of one term of the same quantity. The body of work on redundancy in motor control is not displaced by the reading proposed here: it supplies one of the two required ingredients, and what was missing is the second, the behavior of overlap as subset size varies.

Using these measures finally rests on three assumptions that neuromechanical models do not satisfy as they stand, on the nature of the ensemble observed, on the estimation of the entropies, and on the way the noise is injected; section 6 draws the corresponding adaptations.

## 3. A different reading of earlier work

The empirical data exist, in some cases for twenty years; what was missing was a reading able to decide what they mean. Three studies illustrate this, each meeting part of the three conditions set out in the introduction, none of them meeting all three.

The screen of the crustacean pyloric circuit by Prinz and colleagues varied seven synaptic strengths and the intrinsic properties of three neuron classes independently, producing about twenty million circuits, each simulated and then scored against fifteen criteria drawn from biological recordings (Prinz et al., 2004). About two percent satisfied all of them. The striking result is not that percentage but the range of the parameters involved, two orders of magnitude of synaptic conductance. This establishes contribution overlap, the first of the two conditions set out in section 2 and the one equation (2) measures, but not the second. Each circuit was simulated in a single condition, so the protocol made it impossible to check whether two isofunctional circuits would remain isofunctional under a perturbation. The published data already carry the trace of what is missing: while most synapses vary freely, one of them stays systematically weak in the functional circuits. Equation (4) is built to detect this kind of asymmetry, which indicates that the contributions to the output are not interchangeable. Current exploration algorithms run into the same limit in their behavior maps. The result was nonetheless read as a biological confirmation, not as an indication of the nature of the solution space.

The second case (Powell et al., 2021) studies another biological network of the crab stomatogastric ganglion, the gastric mill circuit. Its rhythm can be triggered either by stimulating a projection neuron or by applying a peptide. Although isofunctional (the rhythms produced are comparable), the two pathways recruit distinct neurons and distinct mechanisms. This architecture matches exactly what our reading sets out to test.

To test these circuits, the authors applied two perturbations, a peptide hormone and a proprioceptive afferent. The hormone separates the two rhythms. The strength of the result rests on a strict demonstration: using dynamic clamp to manipulate conductance artificially, the authors showed that the hormone targets exactly the same ionic current in the same neuron in both cases. An identical perturbation therefore produces divergent outcomes, which proves that the generating mechanisms differ. The conditions for degeneracy (isofunctionality, control of the target, divergence under perturbation) are fully met here, and the authors use the term themselves. The sensory input produces similar effects on both circuits. Far from being a failure, that result (discussed in section 4) shows that divergence revealed by one perturbation does not rule out convergence under another, if the second does not target the mechanisms that differ.

Unlike the isolated circuits above, which have neither biomechanics nor proprioceptive feedback, the third study (Cattaert et al., 2026) couples a spinal network (12 neurons, 30 synapses) to a two-muscle musculoskeletal model. The network is driven by simple step commands, which set its state and initiate the movement without supplying any dynamic information. All temporal structuring of muscle activity therefore emerges from the neuromechanical interaction alone. The aim was to determine whether the triphasic pattern of fast movements is imposed by supraspinal commands or generated autonomously by the spinal cord. By reproducing the pattern from static inputs, the authors confirm that the spinal network generates it on its own. To test the mechanism, the topology of the complete model was then compared with four lesioned versions (ablation of an afferent population, of a class of interneurons, or of the crossed flexor-extensor connections). Table 2 summarizes what each one retains.

Spindle feedback is necessary for any valid movement, tendon organ feedback is not, since minimum-jerk trajectories survive its removal. That same tendon feedback is necessary for the triphasic pattern. These reduced models amount to looking for the minimal circuit able to produce the movement, and the answer depends on what producing the movement means: under a kinematic definition of the output, a strongly reduced circuit is enough; under the triphasic definition, the complete circuit is required. The same system therefore has different minimal structures depending on what is retained as its function, which returns to the problem posed in section 2: isofunctionality holds only relative to an output, and that output is chosen by the observer, not dictated by the system.

Table 2: Capacities of the five topologies tested by Cattaert et al. (2026), under two definitions of the output: the production of valid minimum-jerk movements, and the presence of the triphasic pattern.

| Network | Valid minimum-jerk movements | Triphasic pattern |
|---|---|---|
| Complete (Ia + Ib + all interneurons) | full behavioral domain | present, fast and large movements |
| No Ib feedback, crossed connections retained | reduced domain; slower movements | never observed |
| Ia only, no crossed connections | strongly reduced domain | never observed |
| No Ia feedback | none | — |
| All interneurons, no proprioceptive feedback | none | — |

At a finer grain still, the same study replays a given movement with identical parameters while neutralizing each of the four afferent pathways in turn, then removing a single synapse at a time across five configurations, neutralizing meaning holding the membrane potential constant rather than taking the element out of the network. This is the element-by-element intervention our measures require. It served to establish the mechanism of the triphasic pattern, and the generalization rests on 5,670 valid movements. The diversity of descending commands found by the exploration algorithm is attributed there to the redundancy of parallel sensorimotor pathways, whereas it meets the requirement for degeneracy set out above. That diversity appeared as a by-product of the demonstration rather than as its object, which explains why it was not examined as such. Yet it is that demonstration which makes our question possible: one does not ask about the repertoire of a system one doubts can produce anything on its own.

The three cases thus share out the three conditions set out in the introduction without any one of them meeting all three. The pyloric screen establishes the multiplicity of solutions at scale but applies no perturbation. The gastric mill study meets two of them, but on an isolated circuit, without a body and without afferent feedback. The neuromechanical model supplies the closed loop and the peripheral apparatus the other two lack, along with the possibility of intervening on any element of the circuit, but it was built to establish something else. The gap is therefore precise: applying the requirement for degeneracy to a system in which the neural circuit and the musculoskeletal apparatus are coupled by proprioceptive feedback.

## 4. How degeneracy and redundancy are distributed along the sensorimotor hierarchy

The expectation announced in the introduction is that the two properties fall on either side of the boundary between the nervous system and the body: the central nervous system is heterogeneous and shaped by selection, so degeneracy would be its signature; the musculoskeletal apparatus is mechanical, so redundancy in the engineering sense would be its own. The expectation fails in three ways. It fails on the neural side, because neural circuits are not uniformly degenerate. Tononi and colleagues showed this on a fully interconnected network, whose contribution overlap is high and whose degeneracy is low: its units all influence the output in the same way and none keeps an independent effect (Tononi et al., 1999). The same authors show it for the corpus callosum. Sectioning it was long held to be behaviorally silent (Akelaitis, 1944), until lateralized testing revealed deficits in split-brain patients (Gazzaniga et al., 1962). It fails on the muscular side, because muscle is not a fixed mechanical element. Fiber type composition shifts with the history of use, toward oxidative properties under prolonged low-intensity load and toward glycolytic properties under brief high-intensity load (Pette and Staron, 2001; Schiaffino and Reggiani, 2011). Two motor units that are interchangeable today therefore need not remain so after a period of altered use. This is exactly the divergence section 2 requires, here at the periphery and on a longer timescale. It fails in the formalism as well, and this third objection would be enough on its own: by equation (4), a degenerate circuit is necessarily redundant, so the two properties cannot be separated.

Take the two gastric mill circuits again. At rest they produce the same rhythm, and that output does not change in what follows. What changes from one case to the other is the perturbation used to question them. Under the hormone the two circuits diverge, which establishes that the underlying mechanisms are not the same: they are degenerate. Under the sensory input no divergence appears and nothing distinguishes them: for that perturbation they are redundant. Same circuits, same output, two opposite verdicts. This is not a shortcoming of the data. The question is badly posed as soon as it treats degeneracy as something a system possesses on its own, in the way it possesses a mass. The measures of section 2 always bear on a system relative to an output and under a perturbation, and removing either specification leaves nothing to measure. Degeneracy is real, but it is not attributed to a system taken alone: it is a three-term relation, a system, an output, and a perturbation. It has the status of solubility, which exists only relative to a solvent without being a matter of viewpoint. None of the three terms is given ready-made. The gastric mill shows this for the perturbation, since changing the agent changes the verdict. Section 2 showed it for the system: depending on whether one counts the cell body alone or the presynaptic unit together with its synaptic weight, the same inputs go from redundant to degenerate, with nothing changed in the circuit. Deciding what the system is means deciding at what grain it is described. The third term, the output, is the one this section puts to work.

Moving from property to relation changes what can be looked for. The question is no longer about a location but about a proportion: for a system described at a given grain, a given output and a given perturbation, what is the relative weight of the two terms of equation (4)? A proportion, unlike a boundary, varies continuously. This is the proposal we defend here. The respective contribution of degeneracy and redundancy shifts gradually along the sensorimotor hierarchy, without ever switching at once, and our expectation is that the neuromuscular interface is the region where the proportion changes fastest. What follows makes that expectation plausible rather than establishing it, and section 5 draws from it two predictions that can be tested, on the sign of what the periphery contributes and on what makes that contribution vary.

The reason to expect this is that the body has to report to the nervous system what it has done, a role that falls to the sensory receptors. These receptors are converters: they translate mechanical quantities into signals the nervous system can read. And these converters are themselves neurons: housed in the muscle, they belong to the circuit they inform, and the muscle hosts them without owning them. They are not equivalent to each other either. The muscle spindle is an active sensor whose sensitivity is set from the center, whereas the Golgi tendon organ is a markedly simpler device, and axon caliber marks the priority the system gives to the first. The motor unit escapes this division for another reason. It brings together a motoneuron and the fibers it innervates, inseparable by construction, and part of the diversity of solutions plays out at that scale, since the same level of tension can be reached by recruiting different units from one movement to the next. The word boundary is the wrong one, since a boundary only separates. What is at stake is a zone of dialogue: the alpha command descends, force is produced, the mechanical state comes back up converted, and the fusimotor command sets from the center what will come back up. An anterior cruciate ligament injury illustrates this: the ligament is innervated, and its rupture deprives the nervous system of a source of information at the same time as it alters the mechanics of the knee. One and the same lesion affects both domains at once, which no boundary could produce.

We have just said that the musculoskeletal apparatus was fixed and that muscle was adaptive. Both statements are true, but not over the same observation window. During a movement of a few hundred milliseconds, fiber composition does not change and two motor units able to produce the same force are interchangeable. At that scale the periphery is redundant, and that is the regime biomechanical modeling describes. Over several weeks of training or immobilization, composition changes and those units are no longer interchangeable. At that scale the periphery is degenerate. Nothing changed in the muscle between the two descriptions, only the time window being looked at. This dependence follows from the requirement itself. Asking whether isofunctional elements diverge in another context presupposes an interval, since it takes time for a context to become another. Observe four hundred milliseconds and no slow change can appear; widen to a few minutes and fatigue appears; widen to a few weeks and fiber remodeling appears. The three scales also differ in the origin of what perturbs the system. An afferent lesion or a mechanical load comes from outside and acts immediately. Remodeling comes from the history of the system and spreads over weeks. Fatigue sits between the two, generated by ongoing operation: the system perturbs itself by working. Duration is not a fourth term for all that, it specifies the third. To perturb is always to perturb for a certain time, and the verdict changes with that interval as it changes with the agent.

Motor control calls redundancy the fact that there are more muscles than degrees of freedom to command. Is this redundancy in the sense retained here, that of elements carrying the same information about the output? Anatomy is not enough to answer. Two muscles carry the same information if they are commanded together, and distinct information if they are commanded separately, while their arrangement is identical in both cases. It is the structure of the command that decides, not the number of actuators. Between two agonists, nothing fixes the share of one from the share of the other, so the distribution has to be estimated by optimization; between an agonist and its antagonist, the actions offset each other in the net moment, and there again neither informs about the other. In both cases there are more actuators than necessary, hence infinitely many solutions. But an excess of actuators is not a duplication of information, it is mechanical over-actuation. The place to look is the receptors.

Spindles and tendon organs sample the same mechanical state through different variables, length and velocity for the former, tension for the latter. These variables are not free of one another during a given movement phase, since limb kinematics and the force-length and force-velocity relations tie them together. Knowing one reduces the uncertainty about the other, which is what equation (2) measures. The reference model confirms this (Cattaert et al., 2026): removing tendon feedback does not abolish movement, spindle circuitry is enough to produce minimum-jerk trajectories. But that removal abolishes the triphasic command, which shows that tendon feedback also carries something the spindle does not. Contribution overlap and preserved independent effects, on the same elements, under two definitions of the output: this is what equation (4) detects. The overlap is not constant. During acceleration, tension and lengthening velocity change together and the two afferent pathways say close to the same thing. While a position is held, tension can vary without length moving, and the two signals dissociate. It is not only variable, it is adjustable: the sensitivity of the spindle to length and to velocity is set separately by the static and dynamic fusimotor commands. The nervous system can therefore modify for itself the share of information its two sensors have in common. This consequence has not been measured and would be worth measuring.

Tononi's measures assume a downstream output, one that receives without sending anything back. A sensorimotor system does not work that way: the spinal circuit commands the muscles, the muscles move the limb, and the limb reports its state through the afferents. Where then should the limit be placed between what counts as the system and what counts as its output? The question has no single answer, and that is what makes it useful. Suppose first that the output is muscle activation. The system measured stops at the motoneurons, and the conversion of activation into force and then of force into movement falls outside the perimeter. Two configurations producing different forces therefore count as two different outputs, even if the resulting movement is the same. The body stays in the loop: the afferents keep coming back up, and only the path from force to movement stays outside the measured system. Now suppose the output is the movement, its amplitude and its velocity. The perimeter widens, and the two configurations count this time as a single output. This change of output is formally a coarse-graining of states, an operation known to redistribute information rather than simply reduce it (Varley and Hoel, 2022). The first cut measures the degeneracy of the spinal circuit, the second that of the ensemble formed by the circuit and the body it commands (Figure 1). Their difference isolates what the musculoskeletal apparatus adds on its own. Formally, $X$ does not change from one term to the other, the same blocks are perturbed. What widens is the length of the causal path whose outcome is measured.

$$\Delta D_{\mathrm{N}} = D_{\mathrm{N}}(X\,;O_{\mathrm{movement}}) - D_{\mathrm{N}}(X\,;O_{\mathrm{activation}}) \quad (5)$$

The gap is evaluated at one point of the behavior map and not over the whole map, for a reason set out in section 6. It should be positive, because the passage from forces to movement loses information: several force pairs produce the same net moment, and several net moments produce movements that are indistinguishable within the tolerance retained. Widening the perimeter amounts to making isofunctional a set of solutions that were not.

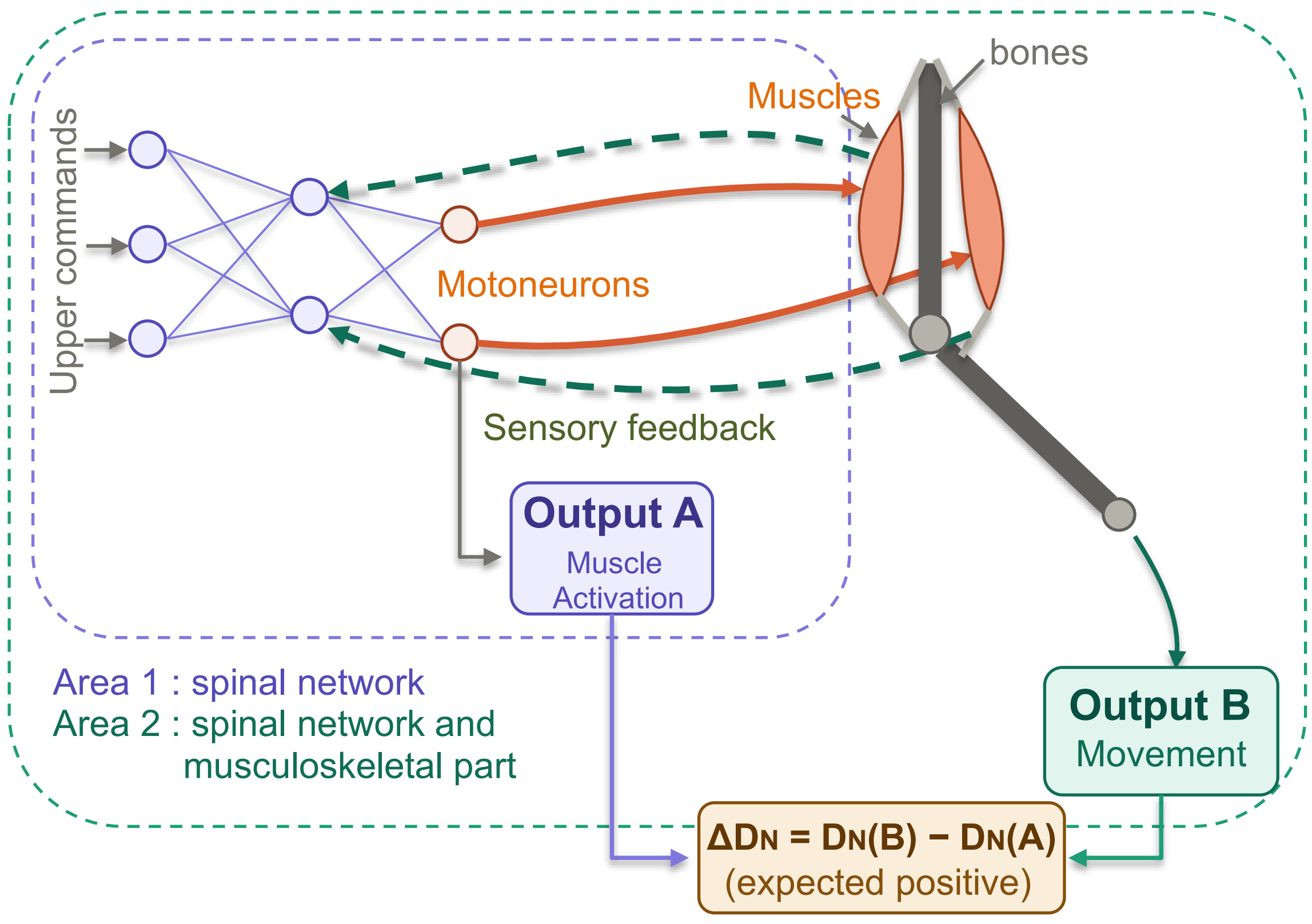

Figure 1: The two definitions of the output and the degeneracy gap. One and the same command configuration passes through the spinal circuit and then through the musculoskeletal apparatus. Output A is the activation, read at the exit of the circuit; output B is the movement produced. Degeneracy is computed twice for the same configuration and the same partition into blocks: only the point of reading changes, which defines two nested perimeters. Their gap, $\Delta D_N$, isolates what the musculoskeletal apparatus adds. The layout is schematic and the construction does not depend on the size of the model.

The model described above offers the barest case in which to carry out this measurement: any mechanical contribution to $\Delta D_N$ can then come only from the distribution of forces between two muscles. Nothing prevents keeping the single joint while increasing the number of muscles that cross it, as at the knee: the underdetermination of actuation increases, the gap keeps the same meaning, and it can be expected to grow. Checking whether it does would be a first extension of this work.

## 5. How the proposal can be tested

The proposal set out above is refutable. We draw five predictions from it, each with what would confirm it and what would refute it, all testable on existing neuromechanical models.

- **The gap between the two measures must be positive**

This is the most direct prediction, since it bears on the measure itself: for the reasons set out in section 4, degeneracy computed with movement as the output must exceed degeneracy computed with muscle activation. Three outcomes are possible. (I) A clearly positive gap confirms the proposal and puts a figure on its scope, by giving the share of degeneracy the body adds to the circuit. (II) A gap near zero refutes it: it would mean that configurations producing the same movement also produce the same forces, that the body transmits without adding anything, and the strict partition we set aside would then be closer to the truth. (III) A negative gap would not be a result but an error, since adding a step that loses information cannot reduce the number of solutions that are merged. This prediction is the least costly to test, because a repertoire of solutions already validated for one behavior supplies the reference configurations: the extra simulations are confined to the neighborhood of each one, provided the noise is applied to subsets of parameters rather than to all of them at once, a condition whose reason section 6 sets out.

- **The gap must grow with the number of muscles**

The previous prediction bears on the sign of the gap, this one on what makes it vary. If the gap measures what the distribution of forces adds, it must depend on the number of actors among which that distribution takes place. Two antagonist muscles offer a single way of varying forces without changing the net moment; a joint crossed by ten muscles offers many more, since each different moment arm multiplies the combinations. The prediction is therefore that the gap grows with the number of muscles, for comparable joints and comparable behaviors. It has an advantage over the previous one: it bears on a variation and not on an absolute value, and the absolute value depends on decisions set out in section 6, the extent of the noise injected and the partition into blocks. Comparing two models is safer than interpreting an isolated figure. An absence of variation would refute the reading we give of the gap, which would then measure something other than the distribution of forces, and what that is would have to be found.

- **Degeneracy must vary with the state of the system**

If degeneracy is a functional relation and not an intrinsic property of anatomy, its local measure must change with the constraints applied. In a sensorimotor system, an internal perturbation such as an afferent lesion removes part of the feedback and redefines the balance of the circuit. It then becomes possible to evaluate the degeneracy of a reference configuration on the intact model, then to repeat the computation for that same configuration facing the lesioned afferent. The comparison bears on the same point of parameter space. Whether the behavior survives is itself a result. A block whose removal leaves the movement unchanged is a block that carried little of what the movement depends on, and the measure taken before the lesion says so in advance, its influence on the movement output having been low however large its influence on activation. Three cases follow. The behavior is conserved, and the two values compare directly, their difference bearing on the lesion alone. The behavior is displaced within the same region, and the comparison holds provided the descriptors stay equally informative in both regimes. The behavior is lost, and there is nothing left to compare, which is the strongest form of the result. Conserved behavior does not mean conserved organization: the same movement can be reached from a circuit whose internal dependencies have been redistributed, and that is where degeneracy is expected to have changed while kinematics shows nothing. The comparison requires in addition that the reduced output be defined the same way in both computations, and that the lesioned block be handled in a declared manner, either removed from the partition on both sides or kept with zero influence. The prediction is that the measured value must change: a configuration showing high degeneracy in the healthy system may see that flexibility collapse under the lesion, which would reflect that the sensory loss has consumed the margins of the network, its overlap. Finding a strictly identical map, whatever the external perturbation, would refute the proposal: it would mean that flexibility does not vary with context, and would reduce degeneracy to a structural constant.

- **The collapse of degeneracy must anticipate breakdown under fatigue**

The periphery looks redundant at the scale of a single movement and degenerate at the scale of adaptation. Slow perturbations must therefore alter internal flexibility before they destroy the movement. Fatigue is the most accessible of those slow perturbations, and its mechanisms reduce to parameters that can be modified in a model: longer latencies and refractory periods on the central side, lower conductivity and loss of contractile capacity on the peripheral side. The prediction here bears on the order of events: if the degeneracy of a solution is evaluated as fatigue degrades it, the measured value must fall gradually while kinematic performance stays preserved. Overlap between the pathways erodes before the breakdown. Under fatigue, two isofunctional solutions will then not necessarily differ in the way their movement declines, but in the order and the pace at which their degeneracy collapses. A degeneracy that stayed stable until the movement broke down would refute this reading of fatigue. Motor physiology already knows a case of this kind: when a submaximal force is held to the point of fatigue, the force stays on target while the electromyographic activity needed to hold it rises, until the task can no longer be maintained.

This prediction has a biological precedent, and it is worth reporting because the effect went unseen at the time. In the crab stomatogastric ganglion, prolonged exposure to elevated extracellular potassium or to unusual temperatures first disorganizes activity, after which the preparation recovers a rhythm close to the one it produced before. Returning to control conditions does not bring it back to the point it occupied in conductance space, and its reaction to the next perturbation is modified (Alonso et al., 2023; Albantakis et al., 2024). These states are called cryptic, precisely because no performance measure taken under control conditions reveals them: macroscopically the preparations are indistinguishable. Two lessons follow for our proposal. A slow perturbation leaves a trace in the organization of the pathways where kinematics sees nothing. And by quantifying internal overlap, the degeneracy index gives a direct reading of that cryptic state: it shows that two solutions indistinguishable today have radically different margins facing the next constraint.

- **Degeneracy must predict which behaviors survive a reduction of the model**

The third prediction asked what the measure becomes after an ablation; this one asks what it announces before. Removing parameters does not cut the reachable domain uniformly: some behaviors disappear, others remain. If degeneracy measures what leaves the system room to vary, it must predict that split. Two lines of reasoning conflict, and that is what makes the prediction discriminating. The first starts from the elements that contribute: a demanding task, close to the limits of the domain, mobilizes the whole circuit, so its degeneracy should be high. The second starts from the solutions available: the more demanding the task, the fewer command sets can satisfy it, so its degeneracy should be low. This second line has a name, the contravariance principle, by which a more constraining objective leaves fewer admissible solutions (Cao and Yamins, 2024). That second line assumes that the number of admissible command sets decides their degeneracy. But number is not what counts, dissimilarity is: a few very different solutions are more degenerate than many nearly identical ones. Only the measurement settles the two. If the first line is right, the behaviors lost are the most degenerate; if the second is right, they are the least.

Each of these predictions can fail without carrying the proposal down with it. One outcome alone would ruin it entirely. If no perturbation, applied anywhere and over any duration, could separate the solutions a model produces for the same behavior, then those solutions would be interchangeable. The second condition would be missing, the observed multiplicity would be redundancy in the strict sense, and the vocabulary the field has used so far would be the right one. We hold that conclusion to be unlikely, given the double dissociation already obtained on a biological preparation, the cryptic states just described, and the fact that reducing a model cuts the reachable domain unevenly.

## 6. How to measure, and under what conditions

The predictions above assume that degeneracy can be computed in a neuromechanical model. It can, but the framework we started from rests on three assumptions that neuromechanical models do not satisfy. We set out the adaptations they force, then what the resulting quantity carries, before situating it against a neighboring framework, complementary to ours in what it can compute.

Table 3: The three assumptions of the original framework that the neuromechanical model does not satisfy.

| | **Original framework (Tononi et al., 1999)** | **Neuromechanical model** |
|---|---|---|
| **Ensemble** | Stochastic process, continuous and stationary (temporal distribution) | Deterministic simulation, transient movement, non-stationary (distribution over configurations) |
| **Entropies** | Linear system with Gaussian distributions (analytic solution from the covariance matrix) | Nonlinear system with non-Gaussian distributions (nonparametric estimators) |
| **Perturbation** | Uncorrelated noise injected continuously, with zero variance on the unperturbed variables | Noise applied from one configuration to the next, around a reference configuration with residual noise |

**What the distribution is over**. An entropy is computed over a set of states with the frequency of each. In the original formulation that set is temporal, which requires a stationary regime. A closed-loop simulation producing a transient movement does not offer one, since its dynamic state changes continuously. Analyzing temporal variation continuously is therefore impracticable. We change the nature of the ensemble accordingly: instead of a system followed over time, we consider a set of configurations, each one a point in the space of the model's free parameters, descending commands and gains of the afferent pathways, together with the activity that results and the movement obtained. The ensemble over configurations then becomes the only valid route, and it has the further advantage of supplying independent draws where successive temporal states would be strongly autocorrelated. This methodological constraint applies only to tasks that fail the stationarity condition: the original temporal analysis remains open in principle for neuromechanical models that meet it.

**How to estimate the entropies**. In Tononi's treatment, the covariance matrix by itself summarizes every departure from independence, which makes the computation analytic. The distribution of the valid configurations of a nonlinear model is not described by those quantities alone. Nonparametric estimators are therefore required, for instance based on distances to nearest neighbors (Kraskov et al., 2004), which assume nothing about its shape and pay for that in data, the number of samples needed growing quickly with the number of dimensions.

**How to perturb**. Since the ensemble is made of configurations, the noise applies to the initial state of each simulation instead of being injected over time. To determine the causal influence of each subset, what matters is that the injected noise be decorrelated from the rest, not that the rest be immobile. In a deterministic simulation the temptation is nonetheless to freeze the other parameters strictly (zero variance), but that freezing raises an obstacle: the output would become a deterministic function of the perturbed block. Two draws close in that block would then be close in the output as well, the distances the estimator relies on would vanish, and the value returned would grow with the number of draws without being bounded by the circuit. To avoid this artifact, one starts from a reference configuration and generates a cloud of draws by applying noise of much larger variance to the block whose influence is being estimated, together with weaker residual noise on the other parameters. This disparity of variance approaches the causal isolation sought while keeping the computation stable.

Two amplitudes are therefore in play, and their ratio governs the result, the information attributed to the evaluated block growing as the residual noise decreases. The ratio must be declared, and held constant across points and across the two definitions of the output.

This need to target the injection of stochastic noise distinguishes the mathematical operation from the physiological perturbations of section 5, such as a mechanical load or an afferent lesion, which apply to the whole system. The distinction might be judged secondary, since a perturbation reaches the whole system anyway through the loop. But propagation is what one undergoes and the entry point is what one chooses. The information-theoretic equations rest on the second: equation (3) calls for the influence of subsets of increasing size, equation (2) for the influence of each block on its own, and a perturbation that varies the entire vector supplies neither of these internal causal dependencies. The mathematical tool and the physiological alteration nonetheless become complementary through comparison. Degeneracy can be evaluated around a given reference configuration, and the same strict computation repeated on an alternative model, lesioned or loaded. This pairwise comparison, feasible over many initial configurations, lets variations in degeneracy serve as an index of the resilience of the system to environmental or structural constraints.

**The choice of blocks**. With about fifty parameters, the enumeration of all subsets required by the original equations, together with the limits of nonparametric estimators in high dimension, is enough to block any computation. The number of elements among which information is distributed has to be reduced, the $n$ of equation (3) becoming the number of blocks. The grouping follows the structure of the circuit modeled, each block gathering the parameters of one pathway. In the model that served as our support, distinguishing the spindle loop, the tendon feedback, the crossed connections, and the descending commands brings some fifty quantities down to a handful of blocks. The constraint comes from the estimator, the decision comes from the architecture of the circuit, and the measure then bears on the share attributable to each pathway, a question the field already asks.

This grouping is not neutral. The formulas, whether they evaluate overlap or derive degeneracy from it, rest entirely on the granularity of the blocks chosen. Encapsulating elements in unbreakable blocks puts the internal dynamics of each pathway, its redundancy and its synergy, beyond the reach of the combinatorial computation, the algorithm measuring only the dependencies between blocks. Changing the boundaries of the blocks therefore changes the value mechanically, and through equation (4) the value of degeneracy, with nothing changed in the underlying circuit. Comparing several partitions then becomes a direct result about the functional organization of the pathways rather than a control. On a smaller system, dispensing with grouping altogether remains the option to prefer.

**Local mapping**. The degeneracy computed this way is local: it characterizes a reference configuration and its neighborhood, not the whole model. The protocol imposes this: one estimates what varying a block around a configuration does to the output, and in a nonlinear system two distant configurations have no reason to share the same sensitivity. Locality organizes the measurement accordingly. A single configuration informs only about its own neighborhood, so one is drawn per region of the behavior map to be characterized. Degeneracy then becomes a field over the parameter domain rather than a number. Nothing leads one to expect that field to be flat: it would be if degeneracy were global, which we have just set aside. The question did not arise in the original framework, where a single network is considered at a time. The local reading therefore does not weaken that framework, it is what the framework becomes when it is applied to a population of parameter sets.

This heterogeneity is instead a useful diagnostic of the partition itself. If the field is homogeneous within a region, the grouping of solutions and the distribution of the informational load agree. A single value per region then becomes a result rather than an assumption. The extent of the noise applied around the reference configuration must be declared with the result, as must the extent of the residual noise applied to the other blocks. A narrow extent linearizes the model locally. A wide extent leaves the region one meant to account for. Between the two, the bound is functional rather than numerical: the draws must stay in the regime of the reference configuration, that is, keep producing a behavior of the same family. Draws that fall outside it estimate the entropies over a mixture of regimes, and the value returned no longer characterizes the configuration one meant to describe.

Nothing above assumes an exploration algorithm. The measurement requires a model that can be replayed, a partition of its parameters into blocks, a reference configuration, and two reduced definitions of the output. The cloud on which the entropies are estimated is not drawn from an existing repertoire, it is generated for the computation. A hand-tuned model, a model calibrated on one participant, or a set of individual models supply as many reference configurations as they have versions. What automatic exploration adds is therefore not the possibility of the computation but the means of choosing where to carry it out: it draws the map over which the reference configurations are distributed, where a single model offers only one.

Finally, applying this measure to isofunctional configurations drawn from distinct families tests the requirement set out in section 2, that of divergence. A redundant repertoire offers solutions that all fail in the same way as soon as the context changes; a degenerate repertoire has some give way and others hold, and it is that difference which leaves the system room to adapt.

**The choice of output descriptors**. At each configuration two quantities are computed, one per definition of the output, and their gap is the $\Delta D_{\mathrm{N}}$ of equation (5). The outputs of the model, whether neuromuscular activation or movement kinematics, are continuous trajectories that no estimator takes as they stand. Each must be reduced to a set of relevant descriptors, for example final position and peak velocity for the movement, or the peaks of the successive bursts of the triphasic pattern for the activation. This reduction is a methodological choice: a global mean or a time integral would smooth the dynamics and fail, since two solutions with opposite time courses could share the same summary. Precise markers discriminate between solutions. Which markers are relevant depends on the question studied and on close knowledge of the model. For the comparison to remain valid, the two reductions must also offer a comparable number of independent dimensions; an imbalance would skew the measured gap mechanically, with nothing in the circuit accounting for it.

**The case of artificial networks**. A framework developed independently for artificial networks approaches this diversity from the other end, quantifying it without assuming anything about its distribution (Huang et al., 2025). It compares independently trained networks pairwise, by their weights, by their activity dynamics, and by the spread of their errors on inputs never met during training. These three quantities are population statistics, none of which has a value for an isolated network, whereas the degeneracy of equation (4) is computed on a single system: the two frameworks stand at the two ends of the axis we have just fixed, which makes them complementary. The third level operationalizes our requirement, since it submits isofunctional solutions to a condition they were not optimized for, and the authors state that it applies to conditions other than theirs: the neuromechanical analogue is immediate, replaying command sets already validated under a load, a modified inertia, a lesioned afferent. The first two levels measure only the extent over which the solutions are spread, that is contribution overlap and not the preservation of independent effects. A measurement program using only those would measure dispersion and call it degeneracy, which would be the same error we started from, in another form.

## 7. What degeneracy implies for motor control

For four decades the central problem of motor control has been posed as a problem of selection. The system has more solutions than the task calls for, the nervous system produces one of them, so there must be a rule that picks it. Successive proposals have put forward candidate rules: minimum jerk, minimum torque change, threshold control, optimal feedback control, muscle synergies. The field has debated which one the nervous system actually uses, and the debate has not converged.

We hold that this is not because the right rule has yet to be found, but because the question itself was produced by the framework in which it was posed. The reasoning takes three steps. If the elements of a redundant system are interchangeable, nothing in the solutions distinguishes them from one another. If nothing distinguishes them, the distinction must come from outside, as a cost or a rule. And if it comes from outside, there can only be one right rule, so the research program necessarily becomes the search for that rule. Each step follows from the one before, and the first rests on the assumption of interchangeability. The selection problem is therefore not an unfortunate residue of the redundancy framework, it is what that framework implies.

In the reading we propose, that assumption does not hold, and the problem goes with it. Degenerate solutions differ in what they do under perturbation, and are therefore already distinguished from one another before any external rule comes in. Asking how the nervous system chooses between equivalent alternatives no longer makes sense, since they are not equivalent. With redundancy, one asked which rule selects a command among equivalent alternatives. With degeneracy, one asks which family of the available repertoire is in use at a given moment, and how the system came to be there. The question bears on the state, the history and the possibilities of the system, not on the optimality of a choice. This is not to say that the field was wrong: it asked the only question its formalism allowed.

It follows that there is no need to decide among the competing theories. Each identifies constraints that are genuinely at work, and each adds an assumption that nothing establishes, that its constraints designate a single point rather than a whole region. Once that assumption is lifted, they become complementary descriptions of different regions of one space. Minimum-jerk formulations capture a kinematic regularity that a large share of the solutions display. Threshold control captures the fact that commands with no temporal structure can nonetheless produce organized antagonist activity, which neuromechanical models reproduce without implementing any equilibrium-point mechanism. Synergy formulations capture a reduction of dimensionality that holds in some regions and not in others.

If the competing theories correspond to identifiable regions, the command sets found by an exploration algorithm should be distributed accordingly. The proposal has already been made in the discussion of a model whose solutions converged with threshold control without implementing anything of the kind (Cattaert et al., 2026). What our reading adds is the reason to expect such a distribution, and the means of telling the resulting families apart.

Two existing formulations point the same way. The uncontrolled manifold hypothesis (UCM; Scholz and Schöner, 1999) and the minimum intervention principle (Todorov and Jordan, 2002) both imply that the system does not select a single point. The UCM shows this most clearly, since it defines the manifold as the set of configurations that leave the performance variable unchanged. But putting equivalence in the definition amounts to never asking whether those configurations are interchangeable or functionally distinct, and that is the question which decides between redundancy and degeneracy.

The idea that the surplus of effectors is abundance rather than redundancy anticipated most of this argument (Latash, 2012). What it lacked was a way of saying why that surplus is a resource. A surplus of interchangeable elements is a stock of spare parts: it protects against loss, and nothing more. A surplus of elements that are not interchangeable is a repertoire: it provides the means of facing conditions that have not yet arisen, precisely because the alternatives will behave differently. Abundance is degeneracy seen from the point of view of the organism rather than the observer, and its warrant no longer rests on an interpretation once the two properties can be told apart.

If solutions are drawn from a repertoire whose use depends on the state of the system, two research habits need revisiting. The first is averaging. The argument that measurements in neuroscience follow a function specific to each individual (Bernard, 2023) also holds for averaging across trials within one subject, and it gains force there. If the solution in use varies from trial to trial with the state of the system, then the variability that averaging erases is not measurement error, it is the structure of the repertoire. Bernstein's observation described that structure, and the analysis methods that followed were largely designed to remove it.

The second concerns discordant results. If a manipulation affects some regions of the repertoire and not others, two studies of the same manipulation can reach opposite conclusions without either being at fault. The debate on the invariance of muscle synergies illustrates this: the synergies extracted during postural control are specific to each participant while remaining stable when the biomechanical context changes (Torres-Oviedo and Ting, 2010), and those of pedaling hold up under varied mechanical constraints (Hug et al., 2011). The gastric mill described in section 4 is such a case, where the verdict changes with the agent used. This is not a replication failure but information about the structure of the repertoire, and it calls for stratification rather than a larger sample.

A related consequence touches the evaluation of neuromechanical models. Common practice is to identify the behavior a model produces best, then to test that behavior. If the property of interest is degeneracy, this practice measures the wrong thing. A model can perform a given task well and resist the perturbations imposed on it for that task without this saying anything about what it does elsewhere. Conversely, a model whose individual solutions resist less well may cover a far wider range of tasks. For an organism facing conditions it has never met, that range is what counts, and an evaluation run on the single best solution does not see it. Our reading suggests another basis for comparison: the number and the diversity of the families a model can deploy, and the range of perturbations under which those families remain distinct. The first term measures the dispersion of the repertoire, the second measures its degeneracy. The second is what decides, the first only gives the scale over which it is read. Together they make a requirement stronger than a behavior map, which says how many behaviors are reachable without saying how many different strategies get there.

## 8. Discussion

The preceding sections have established that the diversity of solutions is a matter of degeneracy, that degeneracy can be measured, and that it generates testable predictions. What remains is to say what we are not claiming, and what the argument does not yet allow us to conclude.

We do not claim that redundancy is absent from sensorimotor systems. The computation of degeneracy requires the opposite, since it rests on a high overlap term, and we have set aside the idea of confining redundancy to the periphery. Nor do we claim that existing work on musculoskeletal redundancy was misdirected: it describes that redundancy where it is most present, and that description is one of the two terms our reading needs. Nor do we claim that our use of the word is the only legitimate one. In mathematics, parameters are called degenerate when they compute exactly the same function whatever the input, and such parameters cannot diverge: our criterion is the reverse of theirs, and we use the word in its biological sense. A third sense circulates in the theory of emergence, where degeneracy refers to the information about the past that is lost when several states converge on the same next state, and counts there as a defect (Varley and Hoel, 2022). Finally, we do not claim that the proposal has been verified on data: the argument is conceptual, and the obstacles to quantifying it, set out in section 6 and completed below, indicate what remains to be done to test it.

The argument has relied on a spinal circuit coupled to a limb with a single joint and two antagonist muscles, driven by signals with no temporal content and operating without gravity. The single joint is an advantage for the comparison proposed, and the measure extends to joints crossed by more muscles. The other simplifications bound what can be concluded. Multi-joint systems introduce biarticular muscles and a coordination in which force feedback might play a particular role, and we do not know whether the distribution we describe holds there. The absence of gravity rules out a regime of continuous postural activity, and with it a family of solutions in which the distribution of forces would probably be expressed differently. The use of neurons without action potentials likewise rules out a source of variability, whose relation to degeneracy has not been examined.

We also do not know how degeneracy is distributed over the behavior map. The fifth prediction of section 5 poses the alternative: degeneracy highest at the limits of the reachable domain, where a demanding task mobilizes the whole circuit, or highest at its center, where constraints are loose and admissible command sets are many. One empirical test already exists and points the second way: in artificial networks, a more demanding task narrows the spread of behaviors and widens the spread of weight sets (Huang et al., 2025), that is, more dissimilar parameters for more similar outputs. Since the circuit here is fixed and only the region targeted changes, the transposition remains an expectation, and this is what the proposed measure would settle.

A last limit lies in the nature of the tool. Unlike mutual information, partial information decomposition (PID) separates shared, unique, and synergistic contributions. It would express degeneracy as the combination of high overlap (isofunctionality) and preserved unique information (divergence under perturbation). Three obstacles block that approach.

Decomposing the terms of equation (4) this way requires moving to *n* sources, hence to the Williams and Beer lattice, whose super-exponential complexity makes each term incomputable. Barrett (2015) shows that his result requires a univariate output and does not extend to the multivariate case, which is what a motor trajectory is. And under the Gaussian approximation the computation becomes tractable again, but all decompositions reduce to a single one, in which overlap equals the smallest of the individual contributions and no longer depends on the correlations between sources. The distinction between redundancy and degeneracy depends on those correlations: the computation goes through, but what was sought is no longer in the result. The 1999 formalism avoids these locks, its terms ($MI^P$) being directly computable and estimable by sampling, without the constrained optimizations of PID. Twenty-five years later, a collective review still refers to it as the measures that quantify degeneracy in biological networks, and recalls the requirement that makes both its strength and its cost: subsets have to be considered at every size (Albantakis et al., 2024). It is that requirement whose practical consequences section 6 draws, by replacing the enumeration of subsets with a grouping into blocks. Tononi's group has since developed other measures, which pursue a different aim (Barbosa et al., 2021) but supply a useful precedent: the admissible partitions there are restricted to those that genuinely disintegrate the mechanism, which makes their choice a matter of theory rather than a computational convenience.

The first step is quantification, under the conditions we have set out. The choice of reference configurations becomes a decision in its own right there, and characterizing what exploration algorithms make available would serve beyond this work. Next comes the confrontation with existing theories of motor control: if each describes a region of the domain, the families of solutions should be distributed accordingly, and that distribution can be checked on data already produced. Extending to human data requires a more demanding shift, not of method but of outlook. The requirement is that of double dissociation: a relevant manipulation is not one that degrades everyone's performance, it is one that affects some participants and spares others. A joint injury is a striking example, since it alters sensory information and mechanical properties at once. Once a manipulation has that property, our proposal predicts that the effect will be read in the separation of participants into groups and not in the displacement of a mean. Nothing here calls for new tools: what changes is the question put to the data.

## 9. Conclusion

A quarter of a century ago, Tononi, Sporns and Edelman separated degeneracy from redundancy and observed that cases of degeneracy were systematically filed under redundancy, for want of a framework that could tell them apart. The observation holds today for motor control. The field has observations that meet the requirement for degeneracy. Families of solutions that diverge under perturbation. Afferent pathways that overlap for one function and not for another. Circuits whose reduced versions succeed under one definition of the output and fail under another. And it describes them in a vocabulary that cannot represent what they show.

That vocabulary has an effect beyond description. Redundancy names a property of the mechanical system, that of having more degrees of freedom than the task requires. In adopting it, motor control received its question ready-made from a neighboring discipline: selecting among solutions already declared equivalent. Biomechanics and motor control have long wanted to work together, and they manage it on data. But a field that has to borrow the other's terms to state its own problem stays beside it, never with it.

Degeneracy changes this, because it can be read neither in the mechanics alone nor in the circuit alone. It requires knowing whether different command sets remain equivalent when conditions change, which means following the effect of the circuit through to the movement and the return of the movement back to the circuit. That is exactly what neuromechanical models make possible, and it is why the question could not be posed before them.

What is at stake is therefore not a word. Redundancy and degeneracy license different questions. With redundancy, one asks how the nervous system chooses between solutions said to be equivalent. With degeneracy, one asks what its repertoire still allows it to do.